\documentclass[12pt]{article}

\usepackage{graphicx}
\usepackage{subcaption}
\usepackage{amsmath, amssymb}
\usepackage{xcolor}
\usepackage{fullpage}
\usepackage{todonotes}
\usepackage{hyperref}
\usepackage[round]{natbib}

\usepackage{algorithm}
\usepackage{algpseudocode}

\renewcommand{\baselinestretch}{1.5}

\newcommand{\sigtx}[1]{\sigma^2_{\text{#1}}}
\newcommand{\tautx}[1]{\tau^2_{\text{#1}}}
\newcommand\independent{\protect\mathpalette{\protect\independenT}{\perp}}
\def\independenT#1#2{\mathrel{\rlap{$#1#2$}\mkern2mu{#1#2}}}

\definecolor{myblue}{HTML}{10B1C3}
\definecolor{myred}{HTML}{C31212}
\title{Meta-Analytics: Tools for Understanding the Statistical
Properties of Sports Metrics
  \protect\thanks{\tiny Alexander M. Franks is a Moore/Sloan Data Science
    and WRF Innovation in Data Science Postdoctoral Fellow
    (\href{mailto:amfranks@uw.edu}{amfranks@uw.edu}).  Alexander
    D'Amour is a Neyman Visiting Assistant Professor in the Department
    of Statistics at UC Berkeley
    (\href{mailto:alexdamour@berkeley.edu}{alexdamour@berkeley.edu}).
    Daniel Cervone is a Moore-Sloan Data Science Fellow at New York
    University (\href{mailto:dcervone@nyu.edu}{dcervone@nyu.edu}).
    Luke Bornn is an Assistant Professor of Statistics at Simon
    Frasier University.
This work was partially supported 
 by the Washington Research Foundation Fund for Innovation in
 Data-Intensive Discovery, the Moore/Sloan Data Science Environments
 Project at the University of Washington and New York University, U.S. National Science Foundation grants 1461435, by DARPA under Grant No. FA8750-14-2-0117, by ARO under Grant No. W911NF- 15-1-0172, by Amazon, and by NSERC.
 The authors are grateful to Andrew Miller (Department of Computer
 Science, Harvard University), and Kirk Goldsberry for sharing data
 and ideas which contributed to framing of this paper.}} 
\author{Alexander Franks, Alexander D'Amour, Daniel Cervone and Luke Bornn} \date{\today}

\begin{document}

\maketitle

\begin{abstract}
\setlength{\baselineskip}{1.5em}
  In sports, there is a constant effort to improve metrics which
assess player ability, but there has been almost no effort to quantify
and compare existing metrics. Any individual making a management,
coaching, or gambling decision is quickly overwhelmed with hundreds of
statistics.  We address this problem by proposing a set of
``meta-metrics'' which can be used to identify the metrics that
provide the most unique, reliable, and useful information for
decision-makers. Specifically, we develop methods to evalute metrics
based on three criteria: 1) stability: does the metric measure the
same thing over time 2) discrimination: does the metric differentiate
between players and 3) independence: does the metric provide new
information?  Our methods are easy to implement and widely applicable
so they should be of interest to the broader sports community.  We
demonstrate our methods in analyses of both NBA and NHL metrics.  Our
results indicate the most reliable metrics and highlight how they
should be used by sports analysts.  The meta-metrics also provide
useful insights about how to best construct new metrics which provide
independent and reliable information about athletes.
\vfill
\end{abstract}

\section{Introduction}

In sports, as in many other industries and research fields, data
analysis has become an essential ingredient of management. Sports
teams, traditionally run by people with experience playing and/or
coaching, now rely heavily on statistical models to measure player
ability and inform strategy decisions \citep{lewis2004moneyball,
  oliver2004basketball}. Over the years, the quantity, scope, and
sophistication of these models has expanded, reflecting new data
sources, methodological developments, and increasing interest in the
field of sports analytics. Despite their inherent promise, new
developments in sports analytics have created a clutter of
metrics. For example, there are at least three different calculations
of the WAR (``Wins Above Replacement'') metric in baseball
\citep{baumer2015openwar}, all of which have the same hypothetical
estimand. In general, any individual making a management, coaching, or
gambling decision has potentially dozens of metrics at his/her
disposal, but finding the right metrics to support a given decision
can be daunting. We seek to ameliorate this problem by proposing a set
of ``meta-metrics'' that describe which metrics provide the most
unique and reliable information for decision-makers. Our methods are
simple to implement and applicable to any sport so they should be of
broad interest to the sports analytics community.

\begin{figure}[!ht]
 \centering
    \includegraphics[width=0.65\textwidth]{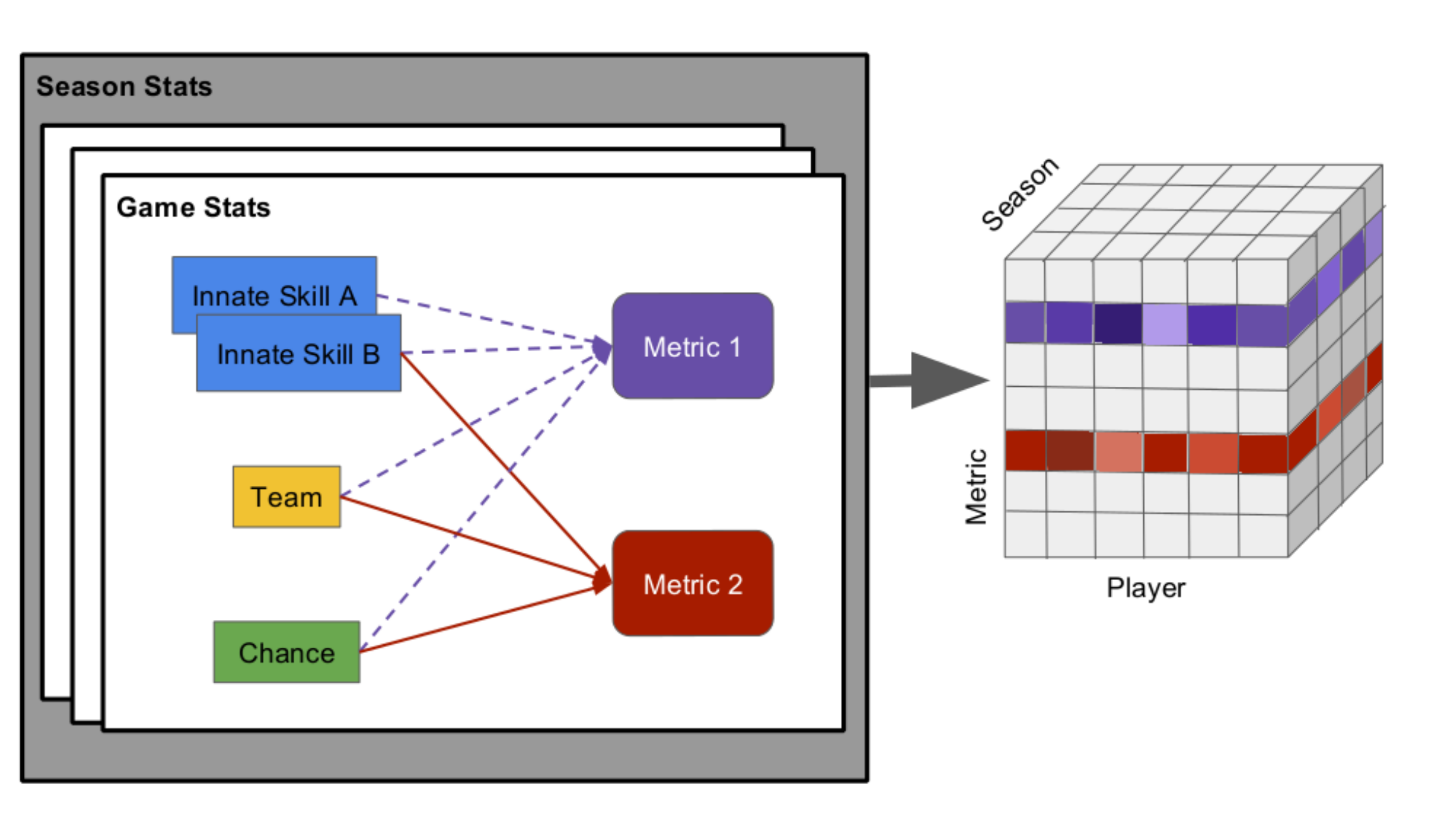}
\caption{Sources of variation in end-of-season metrics.  Player
  metrics confound different aspects of intrinsic player style or
  ability, team effects and chance (e.g. sampling variability).  We
  visualize metrics amongst multiple players across seasons in a
  3-dimensional array (right).  Here, we illustrate two hypothetical metrics, one in
  red and another purple.  Variation in the color's tone on the front
  face corresponds to observed between-player variability in a single season
  and variation on the right face corresponds to variability in the
  metric for one player over time. Team-wise and chance variation also play a
  role in determining the variation in color tone.}
\label{fig:cartoon}
\end{figure}

The core idea of our work is that quantifying sources of
variability---and how these sources are related across metrics, players, and time---is
essential for understanding how sports metrics can be used.  In this
paper, we consider three different sources of variation, which we
classify differently depending on the use-case. These are 1) intrinsic
player skill, 2) context, e.g. influence of teammates, and 3) chance, i.e. sampling
variability. Each of these sources can vary across seasons and
between players.  We consider each player metric to be composed of a
combination of these sources of variation (Figure \ref{fig:cartoon}),
and in this paper we discuss several diagnostics that can be used to
assess how well certain metrics are able to measure, control for, and
average across these sources of variation, depending on what is
required by the decision-maker.

The primary purpose of constructing our meta-metrics is to categorize the
sources of variation in the data as \emph{signal} and \emph{noise}.  The
signal corresponds to variation that is the key input into a decision
process, e.g., a player's ability to operate in a given system,
whereas the \emph{noise} is variation that we choose not to explain
either because of complexity or lack of information (e.g., complex
team interactions or minuscule variations in a player's release
between shots).  When relevant we condition on
observed contextual information (e.g. player position) to create more reliable
and interpretable signals.

For a metric to be useful for a particular decision, its treatment of
variation needs to match up with the decision that is being made.  For
example, consider two distinct tasks in which metrics are often
deployed -- attribution, where we wish to credit a portion of a team's
success to a given player for, e.g., year-end awards, and acquisition,
where we wish to assess whether a player should be added to or
retained on a team.  The classification of signal and noise in these
decision tasks is very different. For attribution, we do not care
whether a player can repeat their performance in another season (or
arguably even how much of their performance was due to chance),
whereas repeatability is a central question in player acquisition.
That is, chance and team context are still relevant signals when
making an attribution decision, but are sources of noise for an
acquisition decision. 

While we can isolate some player-wise, season-wise, and team-wise
variation by subsetting the data, all measurements that we take are
confounded with chance.  Further ``skills'' are abstract concepts
that are often collapsed together.  With this in mind, we define three
meta-metrics that can be used to answer the following questions of
player performance metrics: 

\begin{itemize}
\item \textbf{Discrimination}: Does the metric reliably differentiate between players?
\item \textbf{Stability}: Does the metric measure a quantity which is stable over time?
\item \textbf{Independence}: Does the metric provide new information?
\end{itemize}

Our discrimination meta-metric quantifies how useful a metric is for
distinguishing between players within a given season, whereas our
stability meta-metric measures how much a metric varies season to
season due to changes in context and player skill after removing
chance variation. The independence meta-metric quantifies how much
information in one metric is already captured by a set of other
metrics.  Our meta-metrics are based on ideas which have a long
history in statistics (e.g., analysis of variance) and psychometrics
(e.g., Cronbach's alpha) \citep{fisher1925, cronbach1951coefficient,
  kuder1937theory} but have not received widespread treatment in
sports.  The limited work quantifying the reliability of metrics in
sports mostly appears in blogs \citep{hockeyStabilize, threeStabilize,
  mvpchance} and our hope is to formalize and generalize some of the
ideas discussed in these these articles.  We start, in Section
\ref{sec:methods} by motivating and defining three meta-metrics and
discuss how to estimate them in Section \ref{sec:inference}.  Section
\ref{sec:results} demonstrates the application of these meta-metrics
to player performance in National Basketball Association (NBA) and
National Hockey League (NHL). Lastly, in Section \ref{sec:adjust} we
discuss building new metrics and adjusting existing ones in order to
improve their meta-analytic properties.

\section{Defining Meta-metrics}
\label{sec:methods}

Throughout this paper, we write the 3-dimensional array of players,
seasons and metrics as $X$, with $X_{spm}$ the value of metric $m$
for player $p$ from season $s$ (see Figure \ref{fig:cartoon}).  Our
meta-metrics are all R-squared style statistics and can be understood
as functions of the (co)variances along the three dimensions of $X$.
As a useful example, consider a model for a metric $m$ that varies over time $s$
and between players $p$ is a linear mixed effects model:
\begin{align}
\label{eqn:mixed}
	X_{spm} &= \mu_m + Z_{sm} + Z_{pm} + Z_{spm} + \epsilon_{spm},
\end{align}
where
\begin{align*}
	Z_{sm} & \sim [0, \sigtx{SM}] \\
	Z_{pm} & \sim [0, \sigtx{PM}] \\
	Z_{spm} & \sim [0, \sigtx{SPM}] \\
	\epsilon_{spm} & \sim [0, \tautx{M}],
\end{align*}
and $[\mu, \sigma^2]$ represents a distribution with mean $\mu$ and variance $\sigma^2$.
The terms $Z_{*}$ can be thought of as random effects, while
$\epsilon_{spm}$ represents individual player-season variation in a metric---for instance, binomial variation in made shot percentage given a finite sample size. $Z_{spm}$ and
$\epsilon_{spm}$ are distinguished by assuming that for an infinitely long
season, a player's metric would have no such variability, thus
$\epsilon_{spm} = 0$. Note that we can recognize
$\sigtx{PM} + \sigtx{SPM} + \tautx{M}$ as the within-season,
between-player variance; $\sigtx{SM} + \sigtx{SPM} + \tautx{M}$ as the
within-player, beween-season variance; and of course,
$\sigtx{SM} + \sigtx{PM} + \sigtx{SPM} + \tautx{M}$ as the total
(between player-season) variance. Both the discrimination and stability meta-metrics defined in this section can be expressed as ratios involving these quantities, along with the sampling variance $\tautx{M}$. 

The linear mixed effects model \eqref{eqn:mixed} may be a reasonable choice for some metrics and, due to its simplicity, provides a convenient example to illustrate our meta-metrics. However, an exchangeable, additive model is not appropriate for many of the metrics
we consider.  A major practical challenge in our analysis is that all
of the metrics have unique distributions with distinct support---percentages are constrained to the unit interval, while many per game
or per season statistics are discrete and strictly positive.  Other
advanced metrics like ``plus-minus'' or ``value over replacement''
(VORP) in basketball are continuous real-valued metrics which can be
negative or positive.  

To define meta-metrics with full generality, consider the random variable $X$, which is a single entry $X_{spm}$ chosen randomly from $X$. Randomness in $X$ thus occurs both from sampling the indexes $S, P$, and $M$ of $X$, as well as intrinsic variability in $X_{spm}$ due to finite season lengths. We will then use the notational shorthand
\begin{align*}
	E_{spm}[X] &= E[X | S = s, P=p, M = m] \\
	V_{spm}[X] &= Var[X | S=s, P=p, M=m]
\end{align*}
and analogously for $E_{sm}[X], V_{sm}[X], E_{m}[X]$, etc. For example, $E_{sm}[V_{spm}[X]]$ is the average over all players of the intrinsic variability in $X_{spm}$ for metric $m$ during season $s$, or $\sum_p Var[X_{spm}] / N_{sm}$, where $N_{sm}$ is the number of entries of $X_{s\cdot m}$.

\subsection{Discrimination}
\label{sec:disc}


For a metric measuring player ability to be applicable, it must be a
useful tool for discriminating between different players.  Implicit in
this is that most of the variability between players reflects true
variation in player ability and not chance variation or noise from small sample sizes.  As a useful
baseline for discrimination, we compare the average intrinsic
variability of a metric to the total between player variation in this metric.  A
similar approach which partially inspired this metric was used to
compare how reliably one could differentiate MVP candidates in Major
League Baseball \citep{mvpchance}.


To characterize the discriminative power of a metric, we need to
quantify the fraction of total between player variance that is due to
chance and the fraction that is due to signal.  By the law of total variance, this can be decomposed as
$$ V_{sm}[X] = E_{sm}[V_{spm}[X]] + V_{sm}[E_{spm}[X]].$$
Here, $V_{sm}[X]$ corresponds to the total variation in
metric $m$ between players in season $s$, whereas $E_{sm}[V_{spm}[X]]$ is the
average (across players) sampling variability for metric $m$ in season $s$.  With this decomposition in mind, we
define the discriminative power of a metric $m$ in season $s$ as
\begin{equation}
\label{eqn:disc}
\text{(Discrimination)} \hspace{1cm} \mathcal{D}_{sm} = 1- \frac{E_{sm}[V_{spm}[X]]}{V_{sm}[X]}.
\end{equation}
Intuitively, this describes the fraction (between 0 and 1) of between-player variance in
$m$ (in season $s$) due to true differences in player ability. Discrimination meta-metrics for different seasons can be combined as $\mathcal{D}_m = E_m[\mathcal{D}_{sm}]$.

 It is helpful to understand the discrimination estimand for the linear mixed effects
model defined in Equation \ref{eqn:mixed}.  When this model
holds, $E_{sm}[V_{spm}[X]] = \tautx{M}$, and
$V_{sm}[X] = \sigtx{PM} + \sigtx{SPM} + \tautx{M}$, the between-player variance (equal for all seasons $s$).  Thus, the discrimination meta-metric under the linear mixed effects model is simply
\begin{align}
  \mathcal{D}_m &= 1 - \frac{\tautx{M}}{\sigtx{PM} + \sigtx{SPM} + \tautx{M}}  \label{eqn:disc_mixed} \\
              & = \frac{\sigtx{PM} + \sigtx{SPM}}{\sigtx{PM} + \sigtx{SPM} + \tautx{M}}. \nonumber
\end{align}


\subsection{Stability}
\label{sec:stab}
In addition to discrimination, which is a meta-metric that describes
variation within a single season, it is important to understand how
much an individual player's metric varies from season to season. The notion
of stability is particularly important in sports management
when making decisions about about future acquisitions.  For a stable
metric, we have more confidence that this year's performance will be
predictive of next year's performance.  A metric can be unstable if it
is particularly context dependent (e.g. the player's performance
varies significantly depending on who their teammates are) or if
a players' intrinsic skill set tends to change year to year (e.g. through
offseason practice or injury).

%
%


%

Consequently, we define stability as a metric, which describes how much we
expect a single player metric to vary over time after removing chance
variability.  This metric specifically targets the sensitivity of
a metric to change in context or intrinsic player skill over time.
Mathematically, we define \emph{stability} as:
\begin{equation}
\label{eqn:stab}
\text{(Stability)} \hspace{1cm} \mathcal{S}_m = 1 - \frac{E_m[V_{pm}[X] - V_{spm}[X]]}{V_m[X] - E_m[V_{spm}[X]]},
\end{equation}
with $0 \leq \mathcal{S}_m \leq 1$ (see Appendix for proof).  Here,
$V_{pm}[X]$ is the between-season variability in metric $m$ for player
$p$; thus, the numerator in \eqref{eqn:stab} averages the
between-season variability in metric $m$, minus sampling variance,
over all players. The denominator is the total variation for metric
$m$ minus sampling variance. Again, this metric can be easily
understood under the assumption of an exchangeable linear model
(Equation \ref{eqn:mixed}).:
\begin{align}
	\mathcal{S}_m &= 1 - \frac{\sigtx{SM} + \sigtx{SPM} + \tautx{M} - \tautx{M}}{\sigtx{PM} + \sigtx{SM} + \sigtx{SPM} + \tautx{M} - \tautx{M}} \label{eqn:stab_mixed} \\
 	& = \frac{\sigtx{PM}}{\sigtx{PM} + \sigtx{SM} + \sigtx{SPM}}. \nonumber
\end{align}
This estimand reflects the fraction of total variance (with sampling
variability removed) that is due to within-player changes over time.
If the within player variance is as large as the total variance, then
$\mathcal{S}_m = 0$ whereas if a metric is constant over time, then
$\mathcal{S}_m=1$.



\subsection{Independence}
\label{sec:ind}

When multiple metrics measure similar aspects of a player's ability,
we should not treat these metrics as independent pieces of information.  This is especially
important for decision makers in sports management who use these
metrics to inform decisions.  Accurate assessments of
player ability can only be achieved by appropriately synthesizing
the available information.  As such, we present a method for quantifying the
dependencies between metrics that can help decision makers
make sense of the growing number of data summaries.

For some advanced metrics we know their exact formula in terms of
basic box score statistics, but this is not always the case.  For
instance, it is much more challenging to assess the relationships
between new and complex model based NBA metrics like adjusted plus
minus \citep{adjustedpm}, EPV-Added \citep{cervone2014multiresolution}
and counterpoints \citep{franks2015counterpoints}, which are
model-based metrics that incorporate both game-log and player tracking
data.  Most importantly, as illustrated in Figure \ref{fig:cartoon},
even basic box score statistics that are not functionally related will
be correlated if they measure similar aspects of intrinsic player
skill (e.g., blocks and rebounds in basketball are highly correlated
due to their association with height).

As such, we present a general approach for expressing
dependencies among an arbitrary set of metrics measuring multiple
players' styles and abilities across multiple seasons.  Specifically, we
propose a Gaussian copula model in which the dependencies between metrics
are expressed with a latent multivariate normal distribution.  Assuming we have $M$ metrics of interest, let $Z_{sp}$ be an $M$-vector of metrics for player $p$ during season $s$, and
\begin{align}
\label{eqn:copula}
	Z_{sp} & \stackrel{iid}{\sim} \text{MVN}(0, C)\\
	X_{spm} &= F_m^{-1}[\Phi(Z_{spm})],
\end{align}
\noindent where $C$ is a $M \times M$ correlation matrix, and
$F_m^{-1}$ is the inverse of the CDF for metric $m$.  We define
independence score of a metric $m$ given a condition set of other metrics,
$\mathcal{M}$, as

\begin{equation}
\label{eqn:ind}
\mathcal{I}_{m\mathcal{M}} = \frac{Var \left[ Z_{spm} \mid \{Z_{spq} : q \in \mathcal{M} \} \right]}{Var[Z_{spm}]} = C_{m,m} - C_{m,\mathcal{M}}C_{\mathcal{M},\mathcal{M}}^{-1}C_{\mathcal{M},m}.
\end{equation}

For the latent variables $Z$, this corresponds to one minus the
R-squared for the regression of $Z_m$ on the latent variables $Z_q$ with $q$ in
$\mathcal{M}$.  Metrics for which $\mathcal{I}_{m\mathcal{M}}$ is
small (e.g. for which the R-squared is large) provide little new
information relative to the information in the set of metrics
$\mathcal{M}$.  In contrast, when $\mathcal{I}_{m\mathcal{M}}$ is
large, the metric is nearly independent from the information contained
in $\mathcal{M}$.  Note that 
$\mathcal{I}_{m\mathcal{M}} = 1$ implies that metric $m$ is
independent from all metrics in $\mathcal{M}$.

We also run a principal component analysis (PCA) on $C$ to evaluate
the amount of independent information in a set of metrics.  If
$U\Lambda U^T$ is the eigendecomposition of $C$, with
$\Lambda = \text{diag}(\lambda_1, ... \lambda_M)$ the diagonal matrix
of eigenvalues, then we can interpret
$\mathcal{F}_k = \frac{\sum_1^k \lambda_i}{\sum_1^M \lambda_i}$ as the
fraction of total variance explained by the first $k$ principal
components \citep{Mardia1980}.  When $\mathcal{F}_k$ is large for
small $k$ then there is significant redundancy in the set of metrics,
and thus dimension reduction is possible.

\section{Inference}
\label{sec:inference}
In order to calculate discrimination $\mathcal{D}_m$ and stability
$\mathcal{S}_m$, we need estimates of $V_{spm}[X]]$, $V_{sm}[X]$,
$V_{pm}[X]$ and $V_m[X]$.  Rather than establish a parametric model
for each metric (e.g. the linear mixed effects model
\eqref{eqn:mixed}), we use nonparametric methods to estimate
reliability.  Specifically, to estimate the sampling distribution of
$X$ within each season (e.g., $Var[X_{spm}]$, or equivalently
$V_{spm}[X]$, for all $s$, $p$, $m$), we use the bootstrap
\citep{efron1986bootstrap}.  For each team, we resample (with
replacement) every game played in a season and reconstruct
end-of-season metrics for each player.  We use the sample variance
of these resampled metrics, $\text{BV}[X_{spm}]$, to estimate the
intrinsic variation in each player-season metric $X_{spm}$.  We
estimate $V_{sm}[X]$, $V_{pm}[X]$ and $V_m[X]$ using sample moments.

Thus, assuming $P$ players, our estimator for discrimination is simply
$$\hat{\mathcal{D}}_{sm} = 1 -
\frac{\frac{1}{P}\sum^P_{p=1}\text{BV}[X_{spm}]}{\frac{1}{P}\sum^P_{p=1}
  (X_{spm}-\bar{X}_{s\cdot m})^2}$$
\noindent where $\bar{X}_{s\cdot m}$ is the average of metric $m$ over
the players in season $s$.  Similarly, the stability estimator for a
metric $m$ is
$$\hat{\mathcal{S}}_m = 1 - \frac{\frac{1}{P}\sum^P_{p=1}\frac{1}{S_p}\sum^{S_p}_{s=1}\left[(X_{spm} -
   \bar{X}_{\cdot pm})^2 -
   \text{BV}[X_{spm}]\right]}{\frac{1}{P}\sum^P_{p=1}
 \frac{1}{S}\sum^{S_p}_{p=1}\left[(X_{spm} - \bar{X}_{\cdot\cdot m})^2
   - \text{BV}[X_{spm}]\right]}$$
\noindent where $\bar{X}_{\cdot pm}$ is the mean of metric $m$ for
player $p$ over all seasons, $\bar{X}_{\cdot \cdot m}$ is the total
mean over all player-seasons, and $S_p$ is the number of seasons
played by player $p$.  

All independence meta-metrics are defined as a function of the latent
correlation matrix $C$ from the copula model presented in Equation
\ref{eqn:copula}.  To estimate $C$, we use the semi-parametric
rank-likelihood approach developed by \citet{hoff}.  This method is
appealing because we eschew the need to directly estimate the marginal
density of the metrics, $F_m$.  We fit the model using the R package
\textit{sbgcop} \citep{sbgcop}.  Using this software, we can model the
dependencies for both continous and discrete valued metrics with
missing values. 

In Section \ref{sec:results}, we use $\mathcal{I}_{m\mathcal{M}}$ to
generate ``independence curves'' for different metrics as a function
of the number of statistics in the conditioning set, $\mathcal{M}$.
To create these curves, we use a greedy approach: for each metric $m$
we first estimate the independence score $\mathcal I_{m\mathcal{M}}$
(Equation \ref{eqn:ind}) conditional on the full set of available
metrics $\mathcal{M}$, and then iteratively remove metrics that lead
to the largest increase in independence score (See Algorithm
\ref{indAlg}).


\begin{algorithm}
\centering
\baselineskip=12pt
  \caption{Create independence curves for metric $m$}\label{indAlg}
  \begin{algorithmic}[1]
      \State $\text{IC}_m \gets \text{Vector}(|\mathcal{M}|)$ 
      \State $\mathcal{M}^* \gets \mathcal{M}$
      \For{$i = |\mathcal{M}| \text{ to } 1$}
      \State $\mathcal{I}_{max} \gets 0$
      \State $m_{max} \gets \text{NA}$
      \For{$\tilde{m} \in\mathcal{M}^*$}
        \State $\mathcal{G} \gets \mathcal{M}^* \setminus \{\tilde{m}\}$
        \If{$\mathcal{I}_{m\mathcal{G}} > \mathcal{I}_{max}$}
        \State $\mathcal{I}_{max} \gets \mathcal{I}_{m\mathcal{G}}$
        \State $m_{max} \gets \tilde{m}$
        \EndIf
      \EndFor
      \State $\mathcal{M}^* \gets \mathcal{M}^* \setminus {m_{max}}$
      \State $\text{IC}_m[i] \gets \mathcal{I}_{m\mathcal{M}^*}$
      \EndFor
      \State \Return $\text{IC}_m$
  \end{algorithmic}
\end{algorithm}

\section{Results}
\label{sec:results}

To demonstrate the utility of our meta-metrics, we analyze metrics
from both basketball (NBA) and hockey (NHL), including both
traditional and ``advanced'' (model-derived) metrics.  We gathered data on 70 NBA metrics from all
players and seasons from the year 2000 onwards \citep{bballref}.  We also
gather 40 NHL metrics recorded from the year 2000 onwards
\citep{nhlref}.  Where appropriate, we normalized metrics by minutes
played or possessions played to ameliorate the impact of anomalous events in our data range, such as injuries and work stoppages; this approach sacrifices no generality, since minutes/possessions can also be treated as metrics.  In the appendix we provide a glossary of all of the
metrics evaluated in this paper.  

\subsection{Analysis of NBA Metrics}
\label{sec:nba}
\begin{figure}[!htb]
    \centering
      \includegraphics[width=.8\textwidth]{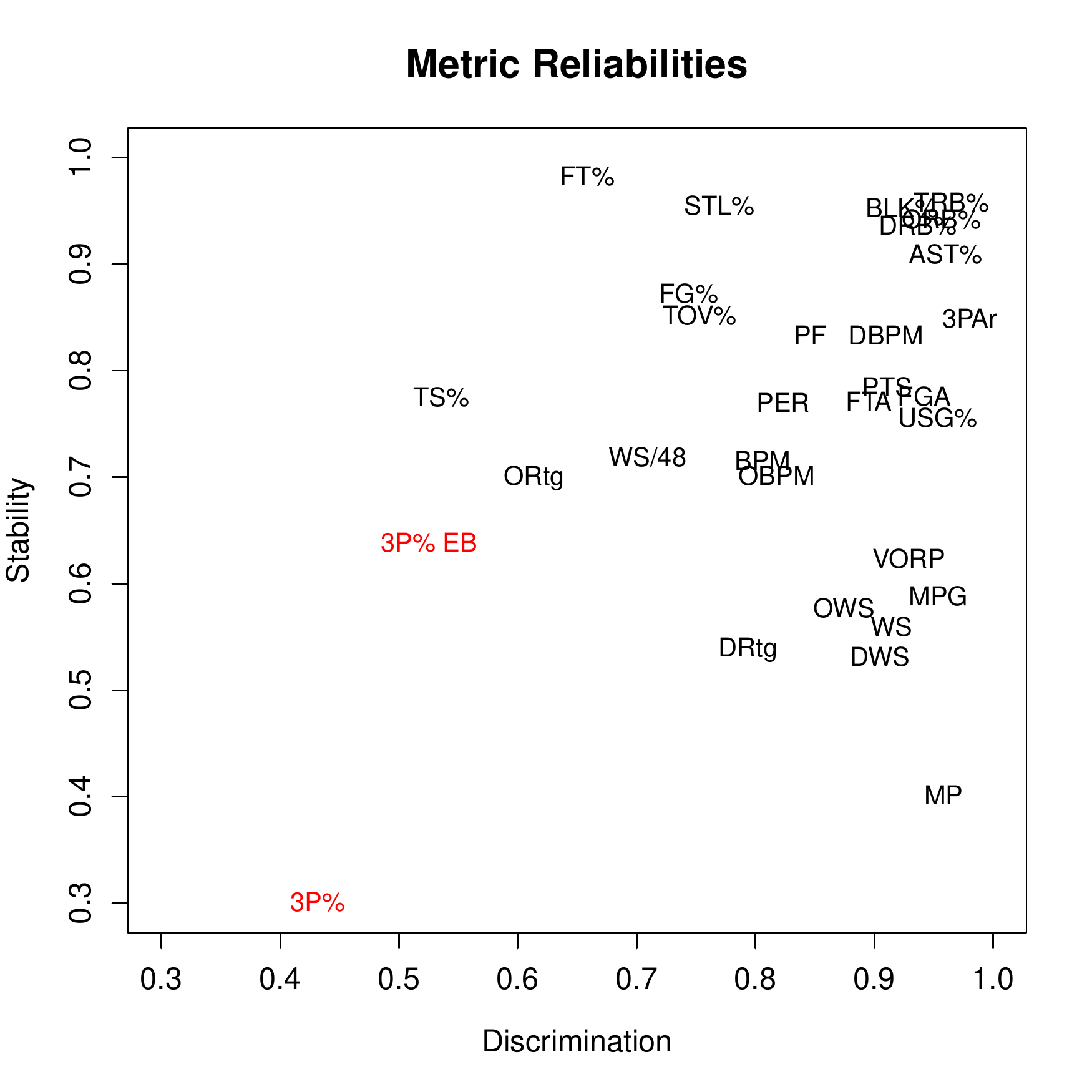}  \\
\caption{Discrimination and stability score estimates for an ensemble
  of metrics and box score statistics in the NBA. Raw three point
  percentage is the least discriminative and stable of the metrics we
  study; empirical Bayes estimates of three point ability (``3P\%
  EB'', Section \ref{sec:adjust}) improve both stability and discrimination .  Metrics like
  rebounds, blocks and assists are strong indicators of player
  position and for this reason are highly discriminative and stable.
  Per-minute or per-game statistics are generally more stable but less
  discriminative.}
\label{fig:nbaRel}
\end{figure}

In Figure \ref{fig:nbaRel} we plot the stability and
discrimination meta-metrics for many of the NBA metrics available on
\url{basketball-reference.com}.  For basic box score statistics,
discrimination and stability scores match intuition.  Metrics like
rebounds, blocks and assists, which are strong indicators of player
position, are highly discriminative and stable because of the
relatively large between player variance. As another example, free
throw percentage is a relatively non-discriminative statistic
within-season but very stable over time.  This makes sense because
free throw shooting requires little athleticism (e.g., does not
change with age or health) and is isolated from larger team strategy and personnel (e.g., teammates do not have an
effect on a player's free throw ability).

Our results also highlight the distinction between pure rate
statistics (e.g., per-game or per-minute metrics) and those that
incorporate total playing time.  Metrics based on total minutes played
are highly discriminative but less stable, whereas per-minute or
per-game metrics are less discriminative but more stable.  One reason
for this is that injuries affect total minutes or games played in a
season, but generally have less effect on per-game or per-minute
metrics. This is an important observation when comparing the most
reliable metrics since it is more meaningful to compare metrics of a
similar type (rate-based vs total). 

WS/48, ORtg, DRtg and BPM metrics are rate-based metrics whereas WS
and VORP based metrics incorporate total minutes played
\citep{bballref}.  WS and VORP are more reliable than the rate based
statistics primarily because MP significantly increases their
reliability, \emph{not} because there is stronger signal about player
ability.  Rate based metrics are more relevant for estimating player
skill whereas total metrics are more relevant for identifying overall
end of season contributions (e.g. for deciding the MVP).  Since these
classes of metrics serve different purposes, in general they should not be
compared directly.  Our results show moderately improved stability and discriminative
power of the BPM-based metrics over other rate-based metrics like
WS/48, ORTg and DRtg.  Similarly, we can see that for the omnibus
metrics which incorporate total minutes played, VORP is more reliable
in both dimensions than total WS.

Perhaps the most striking result is the unreliability of empirical three point
percentage.  It is both the least stable and least discriminative of
the metrics that we evaluate.  Amazingly, over 50\% of the variation
in three point percentage between players in a given season is due to chance.  This is
likely because differences between shooters' true three point shooting
percentage tend to be very small, and as such, chance variation tends
to be the dominant source of variation.  Moreover, contextual
variation like a team's ability to create open shots for a player
affect the stability of three point percentage.

\begin{figure}
  \centering
  \begin{subfigure}{0.4\textwidth}
    \centering
      \includegraphics[width=\textwidth]{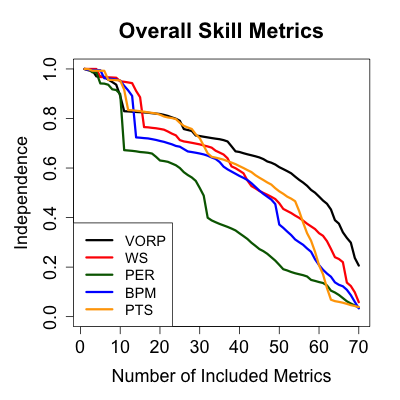}  \\
  \end{subfigure}
  \begin{subfigure}{0.4\textwidth}
    \centering
      \includegraphics[width=\textwidth]{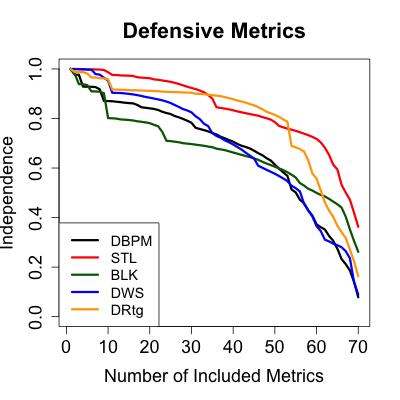}  \\
  \end{subfigure}
\caption{Independence score estimates as a function of the size of the
  conditioning set, for overall skill metrics (left) and defensive
  metrics (right).  The curves look more linear for the overall skill
  metrics, which suggest that they reflect information contained
  in nearly all existing metrics.  The first principal component from
  the five-by-five sub-correlation matrix consisting of the overall
  skill metrics, explains 73\% of the variation.  Defensive metrics
  have independence curves that are more concave.  This highlights
  the fact that defensive metrics are correlated with a smaller
  set of metrics. The first principal component from the five-by-five
  sub-correlation matrix consisting of these defensive metrics,
  explains only 51\% of the variation and the second explains only
  73\%. }
\label{fig:rsq-nba}
\end{figure}

Finally, we use independence meta-metrics to explore the dependencies
between available NBA metrics.  In Figure \ref{fig:rsq-nba} we plot
the independence curves described in Section \ref{sec:inference}. Of
the metrics that we examine, steals (STL) appear to provide some of
the most unique information.  This is evidenced by the fact that the
$\mathcal{I}^{STL}_{\mathcal{M}} \approx 0.40$ , meaning that only
60\% of the variation in steals across player-seasons is explainable
by the other 69 metrics.  Moreover, the independence score estimate
increases quickly as we reduce the size of the conditioning set, which
highlights the relative lack of metrics that measure skills that
correlate with steals.  While the independence curves for defensive
metrics are concave, the independence curves for the omnibus metrics
measuring overall skill are roughly linear.  Because the omnibus
metrics are typically functions of many of the other metrics, they are
partially correlated with many of the metrics in the conditioning set.

\begin{figure}[h!]
    \centering
    \begin{subfigure}[b]{0.3\textwidth}
        \includegraphics[width=\textwidth]{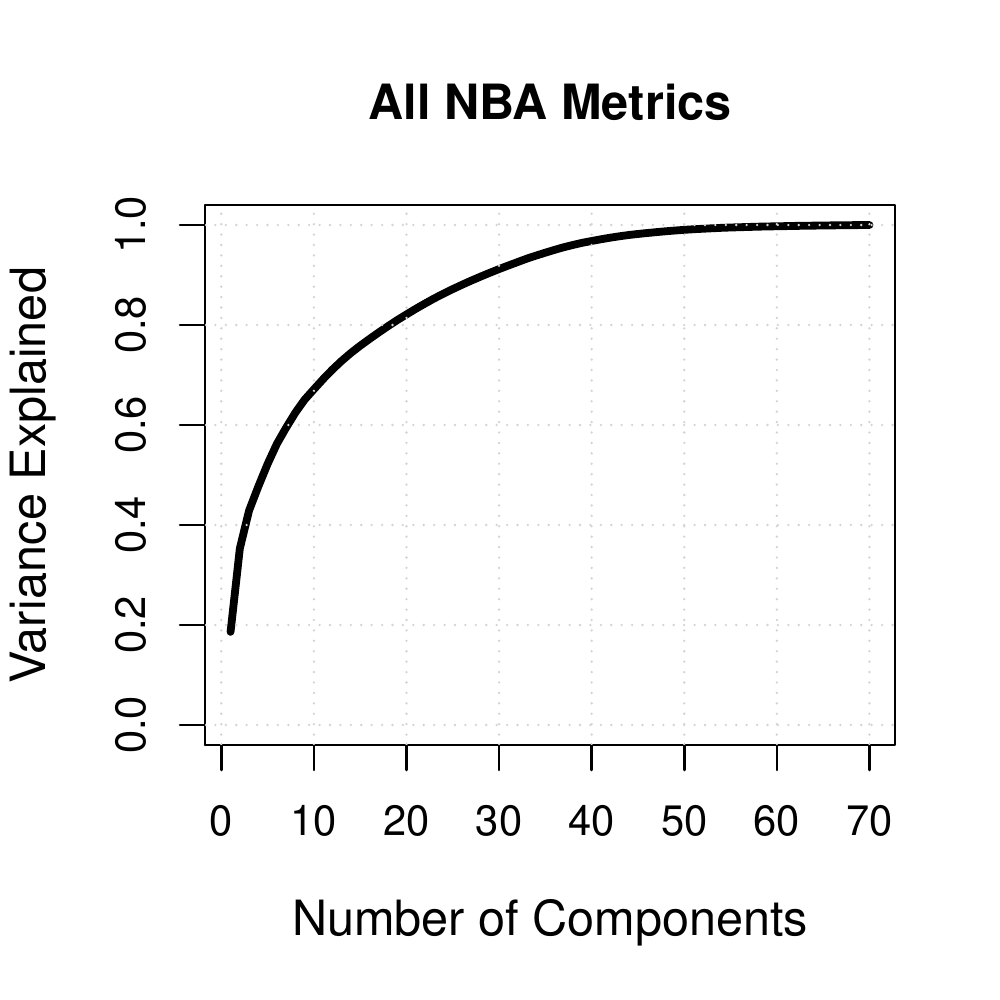}
        \label{fig:nbaVar}
    \end{subfigure}
    \begin{subfigure}[b]{0.3\textwidth}
        \includegraphics[width=\textwidth]{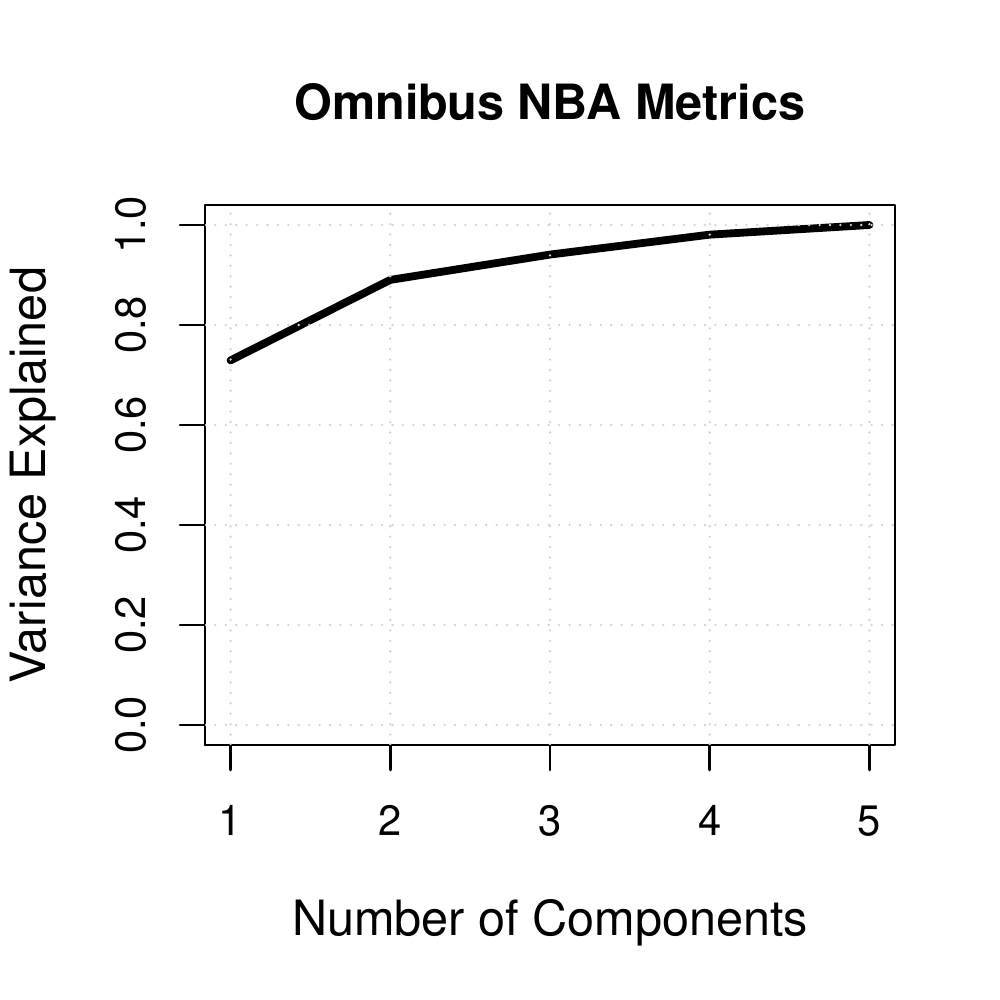}
        \label{fig:omniVar}
      \end{subfigure}
    \begin{subfigure}[b]{0.3\textwidth}
        \includegraphics[width=\textwidth]{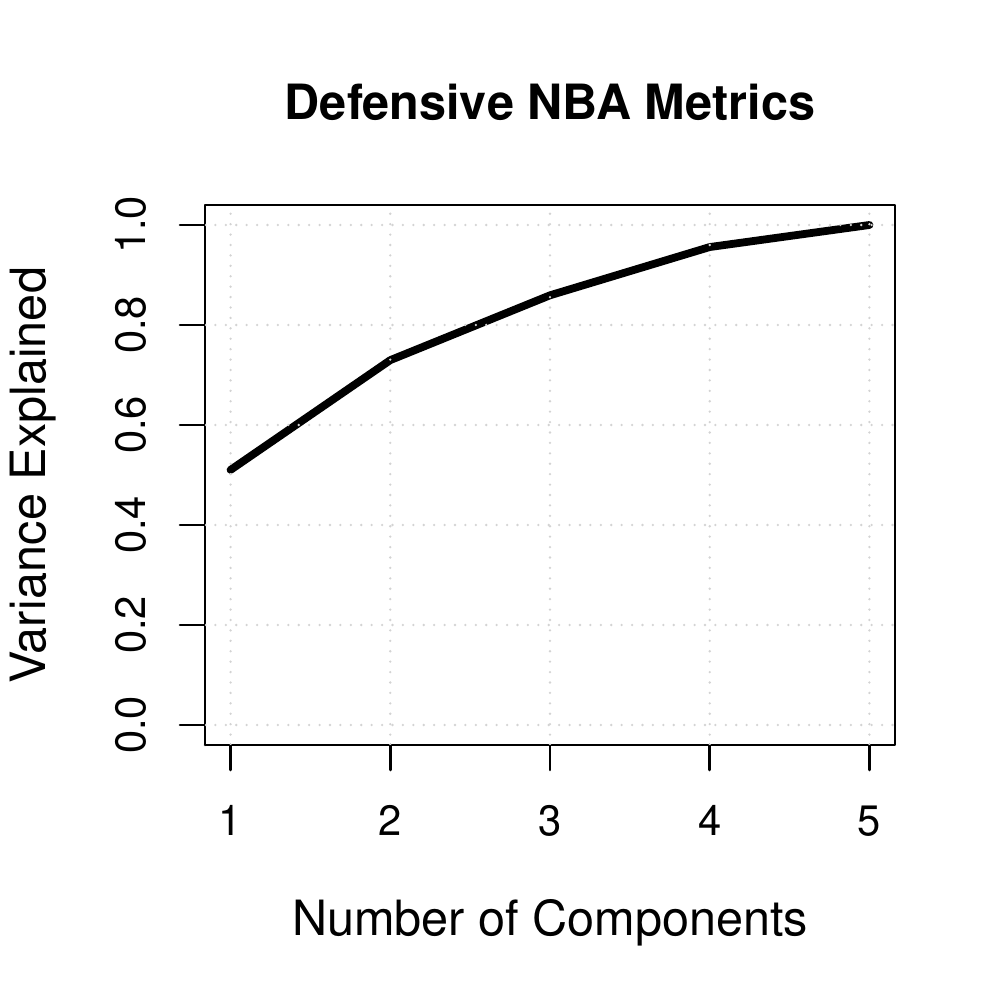}
        \label{fig:defVar}
    \end{subfigure}
\caption{Total variance explained, $F_k$ vs number of principal components
  used.  When evaluating the dependencies among all 70 metrics, we can
  explain over 75\% of the total variability using only 15
  components. For a subset of five omnibus metrics, the first PC
  explains 73\% of the variation, indicating a high level of
  redundancy. For a set of five defensive metrics, the first component
explains 50\% of the variance.}
\label{fig:varExp-nba}
\end{figure}

Not surprisingly, there is a significant amount of redundancy across
available metrics. Principal component analysis (PCA) on the full
correlation matrix $C$ suggests that we can explain over 75\% of the
dependencies in the data using only the first 15 out of 65 principal
components, i.e., $\mathcal{F}_{15} \approx 0.75$. Meanwhile, PCA of the sub-matrix
$C_{\mathcal{M}_o,\mathcal{M}_o}$ where
$\mathcal{M}_o = \{\text{WS, VORP, PER, BPM, PTS}\}$ yields 
$\mathcal{F}_1 = 0.75$, that is, the first component explains 75\% of
the variation in these five metrics.  This means that much of the
information in these 5 metrics can be compressed into a single metric
that reflects the same latent attributes of player skill.  In
contrast, for the defensive metrics presented in Figure
\ref{fig:rsq-nba},
$\mathcal{M}_d = \{\text{DBPM, STL, BLK, DWS, DRtg}\}$, PCA indicated
that the first component explains only 51\% of the variation.  Adding
a second principal component increases the total variance explained to
73\%.  In Figure \ref{fig:varExp-nba} we plot the cumulative variance
explained, $\mathcal{F}_k$ as a function of the number of components
$k$ for all metrics $\mathcal{M}$ and the subsets $\mathcal{M}_o$ and
$\mathcal{M}_d$.

\subsection{Analysis of NHL Metrics}
\label{sec:nhl}
\begin{figure}[!htb]
  \centering
\begin{subfigure}[b]{0.4\textwidth}
    \includegraphics[width=\textwidth]{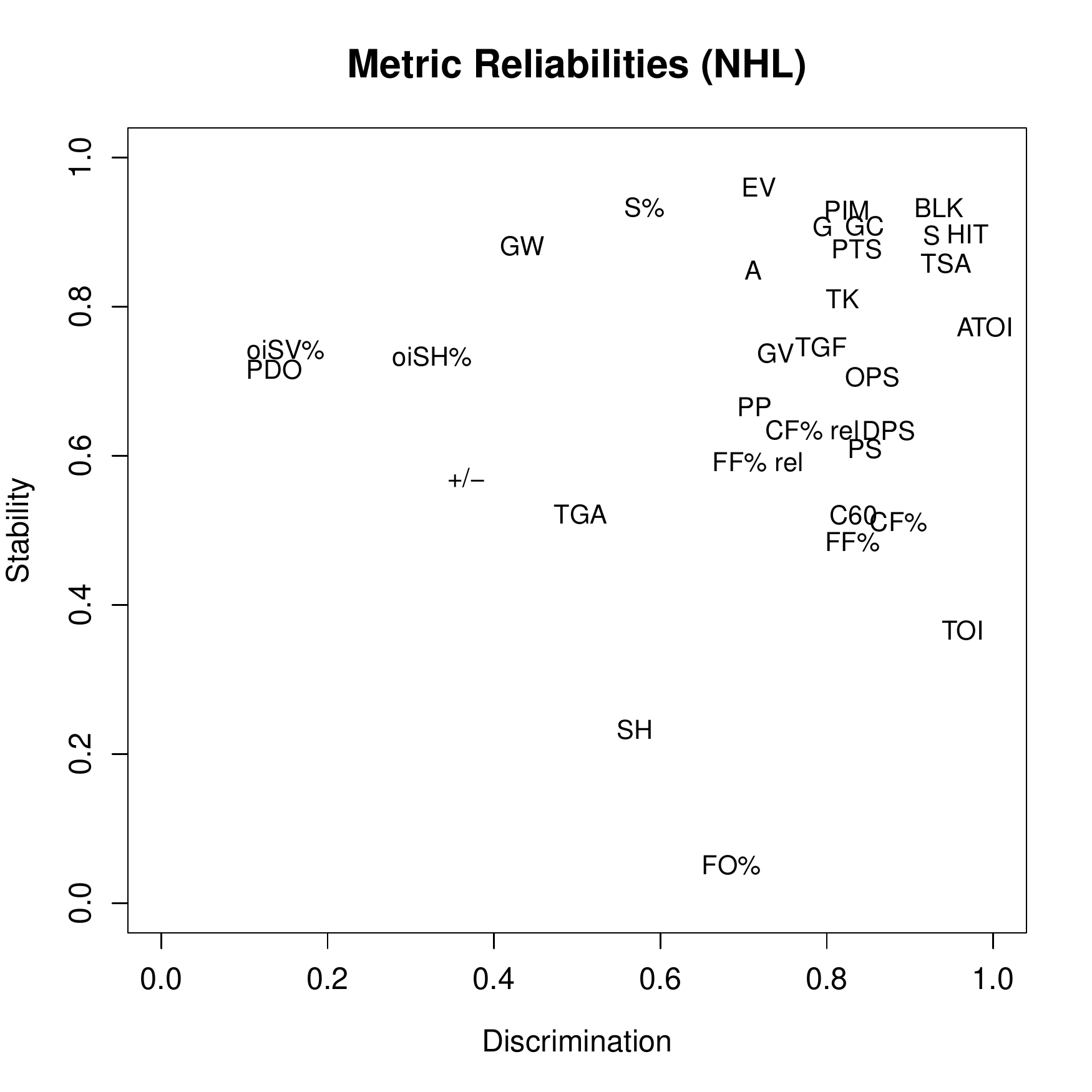}
    \caption{}
    \label{fig:hockeyRel}
\end{subfigure}
\begin{subfigure}[b]{0.4\textwidth}
  \includegraphics[width=\textwidth]{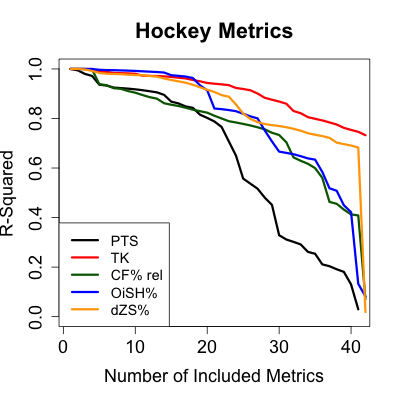}
\caption{}
  \label{fig:hockeyRsq}
\end{subfigure}
\caption{Left) Discrimination and stability scores for many NHL
  metrics.  Corsi-based statistics are slightly more reliable than
  Fenwick statistics.  Plus/minus is non-discriminative in hockey
  because of the paucity of goals scored in a typical game.  Right).  Fraction
  of variance explained (R-squared) for each metric by a set of other
  metrics in our sample.  Only 27\% of the total variance in takeways (TK) is
  explained by all other NHL metrics.   }
\label{fig:hockeyMeta}
\end{figure}

NHL analytics is a much younger field than NBA analytics, and as a
consequence there are fewer available metrics to analyze.  In Figure
\ref{fig:hockeyRel} we plot the estimated discrimination and stability scores for
many of the hockey metrics available on \url{hockey-reference.com}.
Again, we find that metrics like hits (HIT), blocks (BLK) and shots
(S) which are strong indicators for player type are the most
discriminative and stable because of the large between-player
variance. 

Our results can be used to inform several debates in the NHL analytics community.
For example, our results highlight the low discrimination of
plus-minus (``+/-'') in hockey, which can be explained by the relative
paucity of goals scored per game.  For this reason, NHL analysts
typically focus more on shot attempts (including shots on goal, missed
shots and blocked shots).  In this context, it is often debated whether it is better to use
Corsi- or Fenwick-based statistics \citep{CorsiVsFenwick}.  Fenwick-based statistics
incorporate shots and misses whereas Corsi-based statistics additionally
incorporate blocked shots.  Our results indicate that with the
addition of blocks, Corsi metrics (e.g. ``CF\% rel'' and
``CF\%'') are both more reliable and stable than the Fenwick metrics.

In Figure \ref{fig:hockeyRsq} we plot the estimated independence scores as a
function of the number of statistics in the conditional set for five
different metrics.  Like steals in the NBA, we found that takeaways
(TK) provide the most unique information relative to the other 39
metrics. Here, $\mathcal{I}^{TK}_{\mathcal{M}} = 0.73$, meaning that
all other metrics together only explain 27\% of the total variance in
takeaways, which is consistent with the dearth of defensive metrics in
the NHL.  dZS\% is an example of a metric that is highly correlated
with only one other metric in the set of metrics we study, but
poorly predicted by the others.  This metric is almost perfectly
predicted by its counterpart oZS\% and hence
$\mathcal{I}^{dZS}_{\mathcal{M}} \approx 0$ when
$oZS\% \in \mathcal{M}$ and significantly larger otherwise.  This is
clear from the large uptick in the independence score of dZS\% after removing
oZS\% from $\mathcal{M}$.  

\begin{figure}
\centering
\includegraphics[width=.8\textwidth]{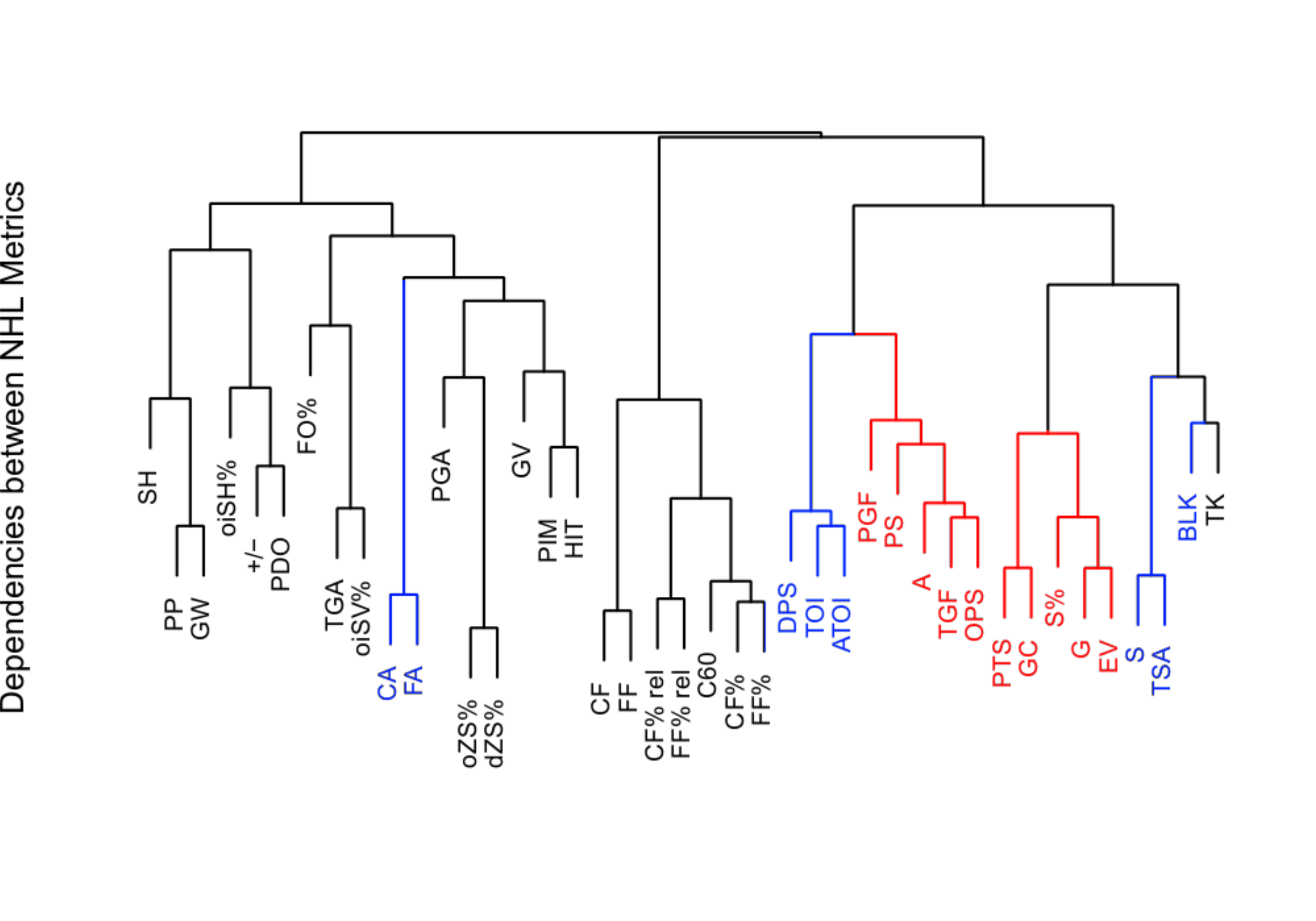}
\caption{Hierarchical clustering of NHL metrics based on the
  correlation matrix, $C$.  Clustered metrics have larger absolute
  correlations but can be positively or negatively associated. The
  metrics that have large loadings on the two different principal
  component (Figure \ref{fig:nhlPCmetrics}) are highlighted in red and
  blue.}
\label{fig:nhlClust} 
\end{figure}

\begin{figure}[h!]
  \centering
  \includegraphics[width=0.5\textwidth]{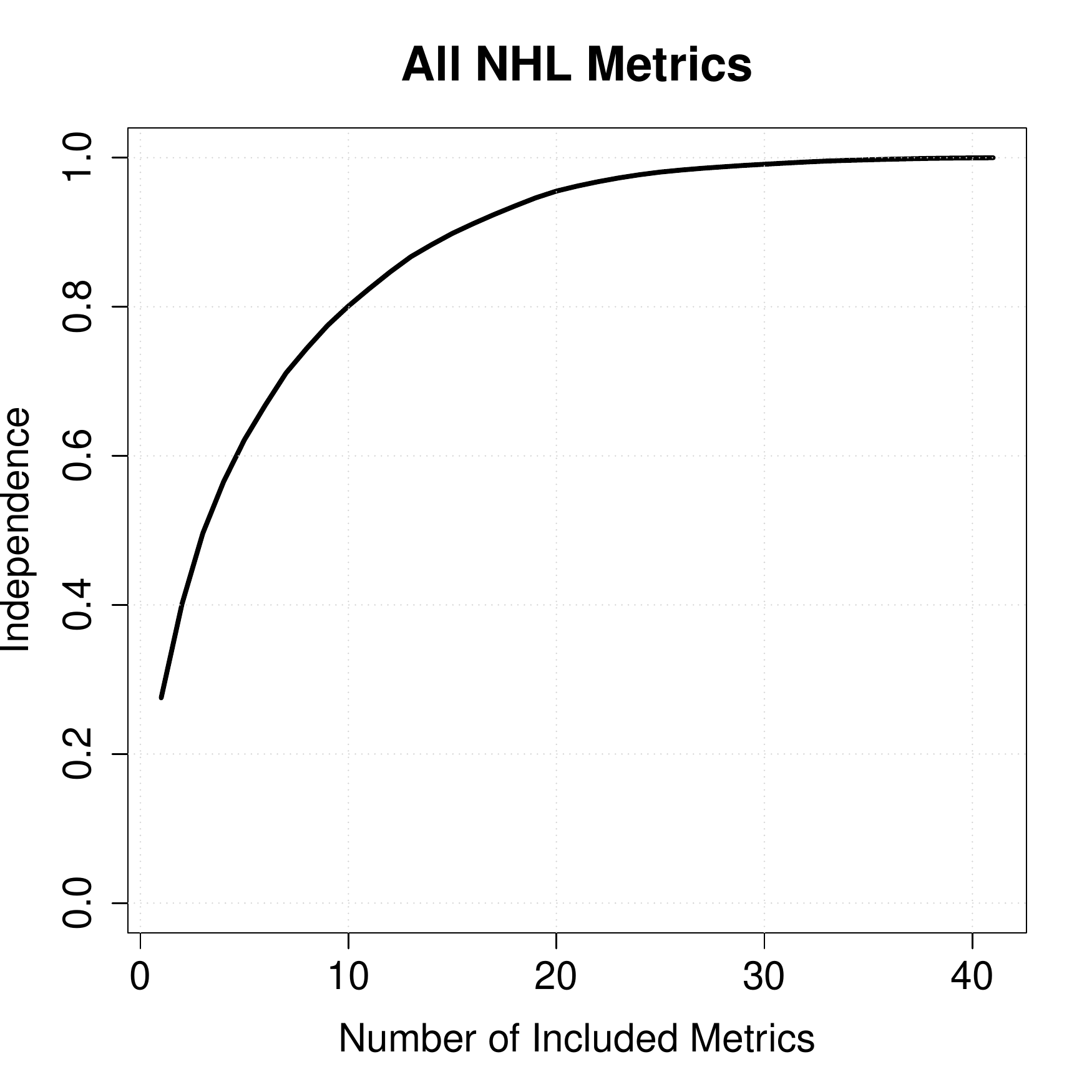}
  \caption{Total variance explained, $F_k$ by number of principal
    components for 40 NHL metrics.  We can
  explain over 90\% of the total variability using only 15
  components.}
\label{fig:varExp-nhl}
\end{figure}

Once again, the analysis of the dependencies among metrics reveals
significant redundancy in information across NHL metrics.  We can
explain over 90\% of the variation in the data using only 15 out of 40
principal components, that is $\mathcal{F}_{15} = 0.90$ (Figure
\ref{fig:varExp-nhl}).  Figure \ref{fig:nhlClust} illustrates a
hierarchical clustering of these metrics based on these dependencies.

\section{Constructing Novel Metrics}
\label{sec:adjust}

In addition to providing useful benchmarks on the quality of different
metrics, the meta-metrics can motivate the design of new and improved
metrics or be used to justify the superiority of new metrics over
traditional ones.  Here we provide two examples in which novel metrics
improve upon existing metrics in at least one of the meta-metrics.  In
the first example, we use a hierarchical model to shrink empirical
estimates of three point ability in basketball.  We demonstrate that
this model-based estimate is both more stable and discriminative than
the simple percentage metric. In the second example, we propose
a method for creating a set of new metrics that are all mutually
independent.

\subsection{Shrinkage Estimators}



Model-based adjustments of common box score statistics can reduce
sampling variability and thus lead to improvements in discrimination
and stability.  In Section \ref{sec:nba}, we showed how three point
percentage was one of the least discriminative and stable metrics in
basketball and thus an improved estimator of three point making
ability is warranted.  We define three point ability using the notation
introduced in Section \ref{sec:methods} as $E_{sp(3P\%)}[X]$ , i.e. the expected three
point percentage for player $p$ in season $s$, and propose a
model-based estimate of this quantity that is both more stable and
discriminative than the observed percentage.

For this model, we assume an independent hierarchical Bernoulli model
for the three point ability of each player:
\begin{align}
\nonumber X^{\text{3P\%}}_{sp} &= \frac{z_{sp}}{n_{sp}}\\
\nonumber z_{sp} &\overset{iid}{\sim} \text{Bin}(n_{sp}, \pi_{sp})\\
\nonumber \pi_{sp} &\overset{iid}{\sim} Beta(r_p \pi^0_p, r_p (1-\pi^0_p)) 
\end{align}
\noindent where $X^{3P\%}_{sp}$ is the observed three point percentage
of player $p$ in season $s$, $\pi_{sp}=E_{sp(3P\%)}[X]$ is the
estimand of interest, $n_{sp}$ is the number of attempts,
$\pi^0_p = E_{p(3P\%)}[X]$ is the career average for player $p$, and
$\pi^0_p(1-\pi^0_p)/r_p$ is the variance in $\pi_{sp}$ over time.  We
use the R package \textit{gbp} for empirical Bayes inference of
$\pi_{sp}$ and $r_p$, which controls the amount of shrinkage
\citep{Kelly}.  In Figure \ref{fig:nbaRel} we plot the original and
shrunken estimates for LeBron James' three point ability over his
career.

We can compute discrimination and stability estimates for the
estimated three point ability derived from this model using the same
approach outlined in Section \ref{sec:inference}.  Although the
empirical Bayes' procedure yields probability intervals for all
estimates, we can still compute the frequentist variability using the
bootstrap (e.g. see \citet{efron2015frequentist}).  In Figure
\ref{fig:nbaRel} we highlight the comparison between observed three
point percentage and the empirical Bayes estimate in red.  Observed
three point percentage is an unbiased estimate of three point ability
but is highly unreliable.  The Bayes estimate is biased for all players,
but theory suggests that the estimates have lower mean squared error
due to a reduction in variance \citep{efron1975data}. The improved
stability and discrimination of the empirical Bayes estimate is
consistent with this fact.

  \begin{figure}
    \centering
      \includegraphics[width=.8\textwidth]{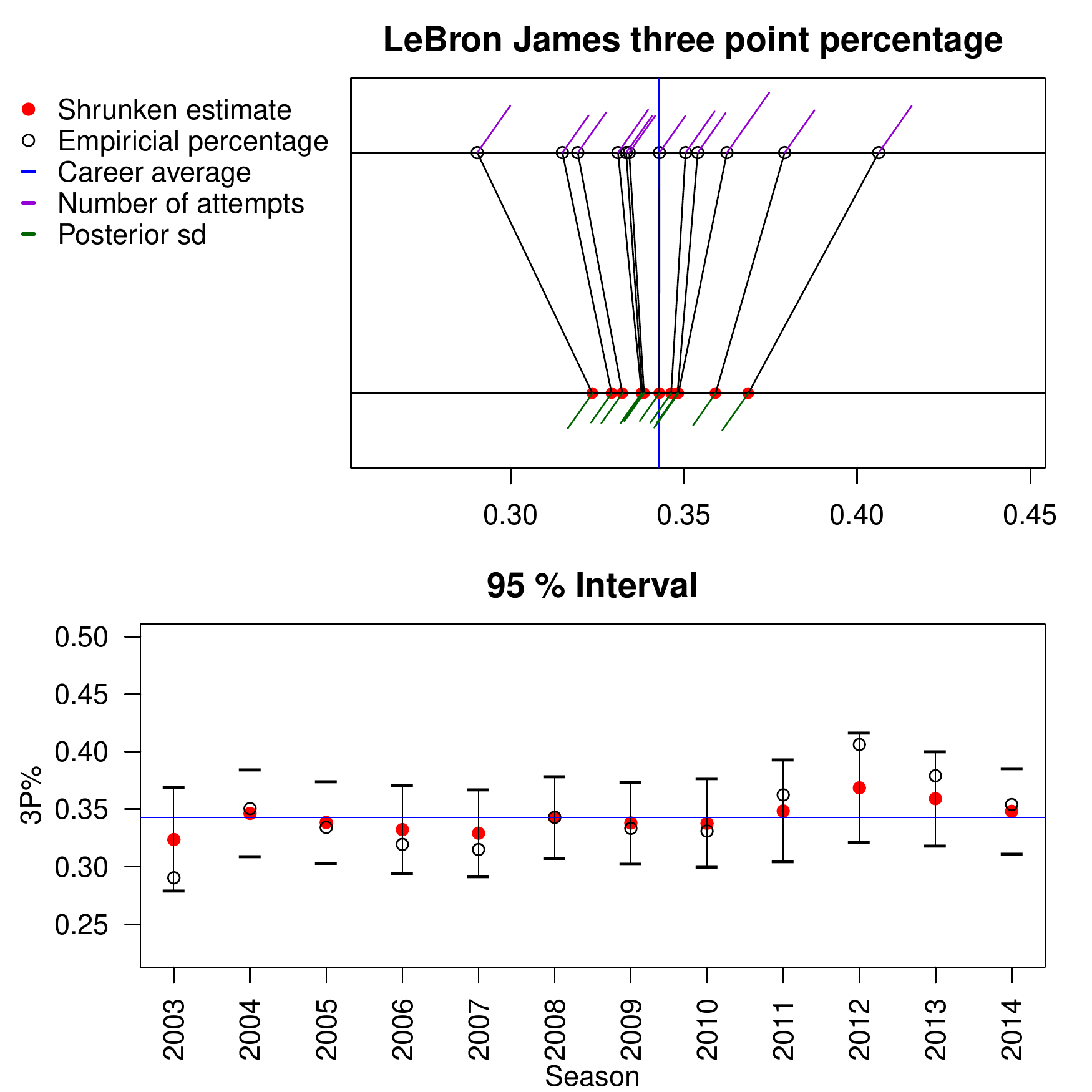}
      \label{fig:bayes}
    \caption{Three point percentages for LeBron James by
  season, and shrunken estimates using the empirical Bayes model
  proposed by \citet{Kelly}.  Shrinking three point percentage to a
  player's career average improves stability and discrimination.}
  \end{figure}

\subsection{Principal Component Metrics}

The dependency model proposed in Section \ref{sec:ind} provides a
natural way to derive new metrics that describe orthogonal aspects of
player ability.  In particular, the eigendecomposition of the latent
correlation matrix, $C$, (Equation \ref{eqn:copula}) can be used to
develop a (smaller) set of new metrics, which, by construction, are
mutually independent and explain much of the variation in the original
set.  If the latent normal variables $Z$ defined in Equation
\ref{eqn:copula} were known, then we could compute the principle
components of this matrix to derive a new set of orthogonal metrics.
The principle components are defined as $W = Z U$ where $U$ is the
matrix of eigenvectors of $C$. Then, by definition,
$W \sim \text{MVN}(0, I)$ and thus
$W_k \independent W_j\ \forall\ k\neq j$.  For the independence score
defined in Section \ref{sec:ind}, this means that
$\mathcal{I}_{k, \mathcal{M}^W_{-k}} = 1$ for all $k$, where
$\mathcal{M}^W_{-k}$ is the set of all metrics $W_j$, $j\neq k$.  We
estimate $Z$ by normalizing $X$, that is
$\hat{Z}_{spm} = \Phi^{-1}(\hat{F}_m(X_{spm}))$ where $\hat{F}_m$ is
the empirical CDF of $X_m$.  Our estimate of the principle components
of the latent matrix $Z$ is then simply
$\hat{W}_{sp} = \hat{Z}_{sp}U$.  

\begin{figure}[!ht]
    \centering
    \includegraphics[width=0.9\textwidth]{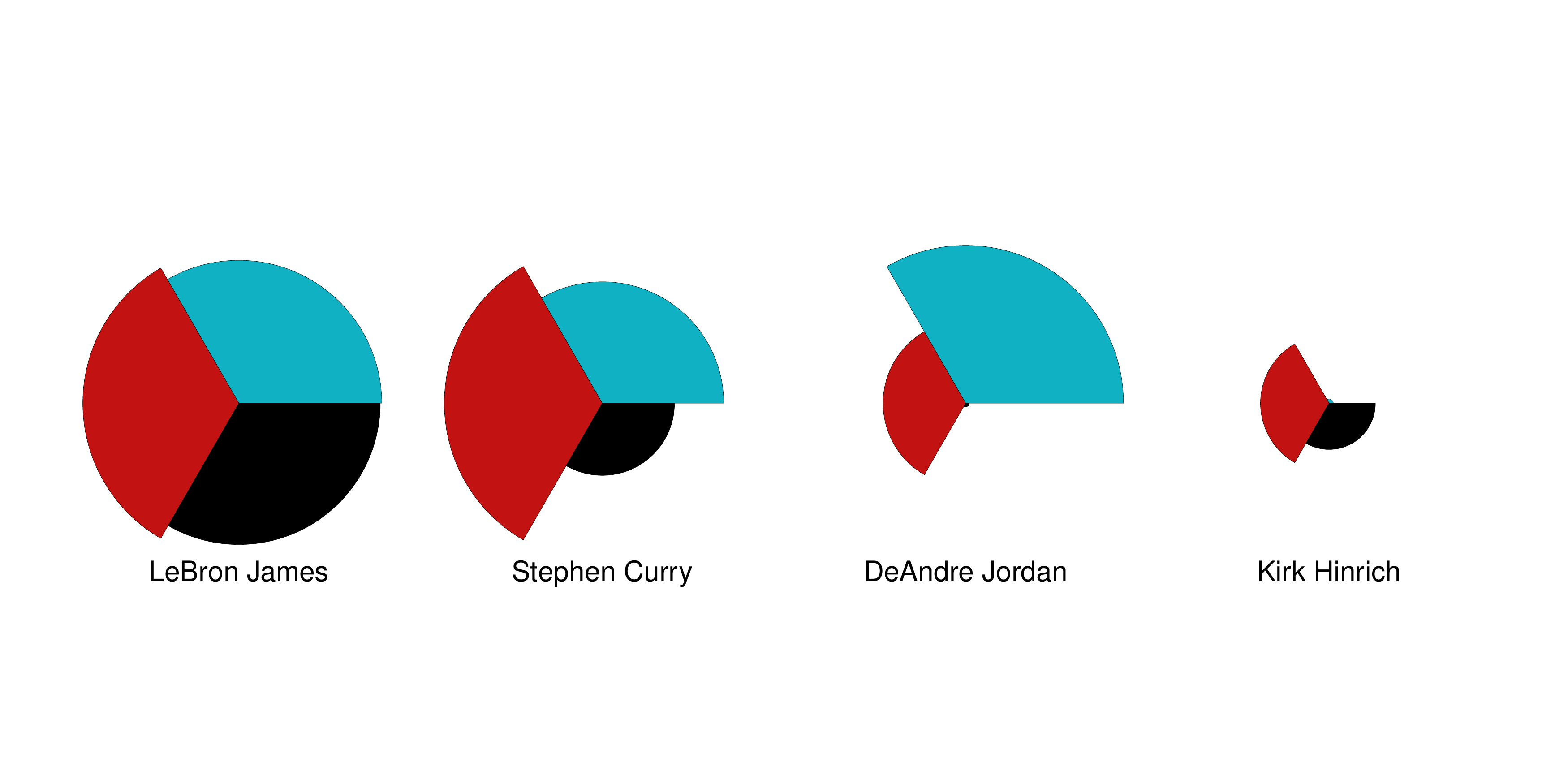}
    \qquad
\raisebox{1.25\height}{
    \begin{minipage}{.3\linewidth}
      \renewcommand{\baselinestretch}{1}
      \footnotesize 
      \centering
      \begin{tabular}{ |c|c|c|}
        \hline
        \multicolumn{3}{|c|}{\textbf{\color{myblue}{``Efficient
        Shooters'' (PC1)}}}\\
        \hline
        \multicolumn{3}{|p{0.75\textwidth}|}{\color{myblue}{ FG\%, PER, WS, \%FG 2P, 2P\%, BPM, TS\%}}\\
        \hline
        Rank & Player & Year \\
        \hline      
        1 & Dwight Howard & 2010 \\
        2 & Dwight Howard & 2009 \\
        3 & Dwight Howard & 2008 \\
        4 & Shaquille O'Neal & 2000 \\
        5 & Shaquille O'Neal & 2004 \\
        6 & Dwight Howard & 2007 \\
        7 & DeAndre Jordan & 2014 \\
        8 & Amar'e Stoudemire & 2007 \\
        9 & Shaquille O'Neal & 2003 \\
        10 & Tim Duncan & 2006 \\
        \hline
      \end{tabular}
    \end{minipage}%
\hspace{.5cm}
    \begin{minipage}{.3\linewidth}
      \renewcommand{\baselinestretch}{1}
      \footnotesize 
      \centering
        \begin{tabular}{ |c|c|c|}
          \hline
          \multicolumn{3}{|c|}{\color{myred}{\textbf{``Shooters, Assisters'' (PC2)}}}\\
          \hline
          \multicolumn{3}{|p{0.75\textwidth}|}{\color{myred}{OBPM, 3PA, AST\%, \%FGA 3P, Avg
          Shot Dist, PGA}} \\
          \hline
          Rank & Player & Year \\
          \hline      
          1 & Stephen Curry & 2014 \\
          2 & Stephen Curry & 2013 \\
          3 & Steve Nash & 2006 \\
          4 & Chris Paul & 2014 \\
          5 & Steve Nash & 2008 \\
          6 & Chris Paul & 2007 \\
          7 & Damon Jones & 2004 \\
          8 & Steve Nash & 2009 \\
          9 & Stephen Curry & 2012 \\
          10 & LeBron James & 2009 \\
          \hline
        \end{tabular}
    \end{minipage}%
\hspace{.5cm}
    \begin{minipage}{.3\linewidth}
      \renewcommand{\baselinestretch}{1}
      \footnotesize 
      \centering
      \begin{tabular}{ |c|c|c|}
        \hline
        \multicolumn{3}{|c|}{\textbf{``High Usage'' (PC3)}}\\
        \hline
        \multicolumn{3}{|p{0.75\textwidth}|}{USG, 2PA, FGA, LostBall, FTA, SfDrawn, PTS, And1}\\
        \hline
        Rank & Player & Year \\
        \hline      
        1 &  Allen Iverson & 2006 \\
        2 & Cory Higgins & 2011 \\
        3 &  Kobe Bryant & 2014 \\
        4 & Allen Iverson  & 2003 \\
        5 & Russell Westbrook & 2014 \\
        6 & Tony Wroten & 2013 \\
        7 & Tony Wroten & 2014 \\
        8 &  Allen Iverson & 2004 \\
        9 &   Jermaine O'Neal & 2004 \\
        10 & Allen Iverson & 2005 \\
        \hline
      \end{tabular}
    \end{minipage}
}
\caption{First three principal components of $C$.  The tables indicate
  the metrics that predominantly load on the components.  Each
  component generally corresponds to interpretable aspects of player
  style and ability.  The table includes the highest ranking players
  across all seasons for each component.  The top row depicts
  principal component score for four players players in the 2014-2015
  season. LeBron James ranks highly among all 3 independent
  components. }
\label{fig:nbaRank}
\end{figure}

We present results based on these new PCA-based metrics for both NBA
and NHL statistics.  In Figure \ref{fig:nbaRank} we list three
PCA-based metrics for the NBA and the corresponding original NBA
metrics which load most heavily onto them. We also rank the top ten
players across seasons according to $\hat{W}_{sp}$ and visualize the
scores for each of these three PCA-based metrics for four different
players in the 2014-2015 season.  Here, the fact that LeBron James
ranks highly in each of these three independent metrics is indicative
of his versatility.  Although the meaning of these metrics can be
harder to determine, they can provide a useful aggregation of
high-dimensional measurements of player skill that facilitate fairer
comparisons of players.

In Figure \ref{fig:nhlPCmetrics} we provide two PCA-based metrics for NHL
statistics.  We again list the metrics that have the highest
loadings on two principal component along with the top ten players (in
any season) by component.  The first principal component largely
reflects variation in offensive skill and easily picks up many of the
offensive greats, including Ovechkin and Crosby.  For comparison, we
include another component, which corresponds to valuable defensive
players who make little offensive contribution. This component loads
positively on defensive point shares (DPS) and blocks (BLK), but
negatively on shots and goals (S, G).

\begin{figure}[ht!]
\centering
\begin{minipage}[t]{.4\textwidth}
  \renewcommand{\baselinestretch}{1}
\vspace{.5in}
     \footnotesize
      \begin{tabular}{ |c|c|c|}
        \hline
        \multicolumn{3}{|c|}{``Offensive skill''}\\
        \hline
        \multicolumn{3}{|p{0.5\textwidth}|}{\textcolor{red}{PTS, OPS, GC, PS, TGF, G,
        A, EV, PGF, TSA}}\\
        \hline
        Rank & Player & Year \\
        \hline      
        1 & Alex Ovechkin & 2010 \\
        2 & Sidney Crosby & 2009 \\
        3 & Alexander Semin & 2008 \\
        4 & Daniel Sedin & 2000 \\
        5 & Evgeni Malkin & 2011 \\
        6 & Daniel Sedin & 2010 \\
        7 & Alex Ovechkin & 2007 \\
        8 & Alex Ovechkin & 2008 \\
        9 & Sidney Crosby & 2012 \\
        10 & Marian Hossa & 2008 \\
        \hline
      \end{tabular}
      \label{fig:nhlPC1}
\end{minipage}
\begin{minipage}[t]{.4\textwidth}
  \renewcommand{\baselinestretch}{1}
\vspace{.5in}
     \footnotesize
      \begin{tabular}{ |c|c|c|}
        \hline
        \multicolumn{3}{|c|}{``Valuable defenders ''}\\
        \hline
        \multicolumn{3}{|p{0.5\textwidth}|}{\textcolor{blue}{ATOI,
        DPS, BLK,\linebreak -S, -TSA, -G, -FA, -CF}}\\
        \hline
        Rank & Player & Year \\
        \hline      
        1 & Nicklas Lidstrom & 2008 \\
        2 & Ryan Suter & 2014 \\
        3 & Toby Enstrom & 2009 \\
        4 & Josh Gorges & 2012 \\
        5 & Toni Lydman & 2011 \\
        6 & Toby Enstrom & 2008 \\
        7 & Chris Progner & 2010 \\
        8 & Paul Martin & 2008 \\
        9 & Niclas Havelid & 2008 \\
        10 & Andy Greene & 2015 \\
        \hline
      \end{tabular}
      \label{fig:nhlPC2}
\end{minipage}
\caption{Player rankings
  based on two principal components.  The first PC is associated with
  offensive ability.  The fact that this is the \emph{first} component
  implies that a disproportionate fraction of the currently available
  hockey metrics measure aspects of offensive ability.  The other included
  component reflects valuable defensive players (large positive
  loadings for defensive point shares and blocks) but players that
  make few offensive contributions (negative loadings for goals and
  shots attempted).  The metrics that load onto these components are
  highlighted in the dendrogram of NHL metrics (Figure \ref{fig:nhlClust}). }
\label{fig:nhlPCmetrics}
\end{figure}

\section{Discussion}

Uncertainty quantification, a hallmark of statistical sciences, has so
far been under-appreciated in sports analytics.  Our work demonstrates
the importance of understanding sources of variation and provides a
method to quantify how different metrics reflect this variation. Specifically, we
explore three different ``meta-metrics'' for evaluating the
reliability of metrics in any sport: discrimination, stability and
independence.
%
Our results show that we can use meta-metrics to characterize the
most discriminative and stable summaries amongst a set of omnibus metrics (like win shares, BPM
and PER for the NBA), which can in turn help decision-makers identify
the metrics that are most relevant for a given task.  Meta-metrics
can also be used as a benchmark for evaluating the improvement of new
estimators.  For instance, in the case of three point percentage, we 
demonstrate that an estimate based on a simple hierarchical model can
improve the stability \emph{and} discrimination of standard boxscore
statistics.  

In this paper, we focused on reliability and dependence of metrics for
\emph{all players in the league} but the meta-metrics can easily be
recalculated for relevant subsets of players.  This is important
because, as shown, in this context the most reliable metrics are often
the metrics which distinguish between player types (e.g., blocks and
rebounds in basketball).  This may be irrelevant when making decisions
involving a specific group of players (e.g., which NBA center to
acquire).  When using metrics to evaluate players of a certain type,
we should compute the meta-metrics conditional on this player type.
For instance, there is less variation in the number of blocks and
rebounds by NBA centers, and as such, these metrics are less
discriminative and stable than they are for the league as a whole.
Moreover, the dependence between blocks and rebounds is largely driven
by height, and thus the conditional dependence between blocks and
rebounds given height is much smaller.  Thus, it is important that the
meta-metrics are always interpreted in the context of the appropriate
group of players.  In light of this point, it is notable that the
meta-metrics that we present in this paper are stated in terms of
expectations and variances, so that estimation of conditional meta-metrics
simply requires replacing marginal expectations and variances with
their conditional counterparts.




Another important consideration is that our meta-metrics only measure
the internal quality of a metric.  The meta-metrics are not designed
to provide any information about how relevant the metrics are for the
sport of interest.  For instance, although we identified Corsi-based
metrics as more discriminative and stable than the related
Fenwick-based metrics, it is still possible that Fenwick metrics are
more predictive of team performance.  As a more extreme example, an
athlete's birthplace zip code would be perfectly discriminative,
stable and independent from all other metrics, but is clearly
irrelevant for determining a player's value to the team.  This
suggests that in practice coaches and analysts should consider a
fourth meta-metric: ``relevance''.  Relevance could simply be a
qualitative description of the metric's meaning or it could a
quantitative summary of the causal or predictive relationship between
the metric and an outcome of interest, like wins or revenue generated.
Nevertheless, the methods presented here provide a useful
characterization of the reliability of existing metrics.  We believe
that future iterations of the meta-metrics outlined in this paper can
become a standard analytical tool that will improve the decisions made
and information gleaned from new and old metrics alike.

\clearpage

\bibliographystyle{natbib}
\bibliography{refs}

\clearpage

\section*{Appendix}

\begin{figure}[h!]
  \caption{Hierarchical clustering of NBA metrics based on the
    correlation matrix, $C$.  Clustered metrics have larger absolute
    correlations (e.g. can be positively or negatively related)}
\label{fig:nbaClust}
  \centering
  \includegraphics[width=1.1\textwidth, angle=-90]{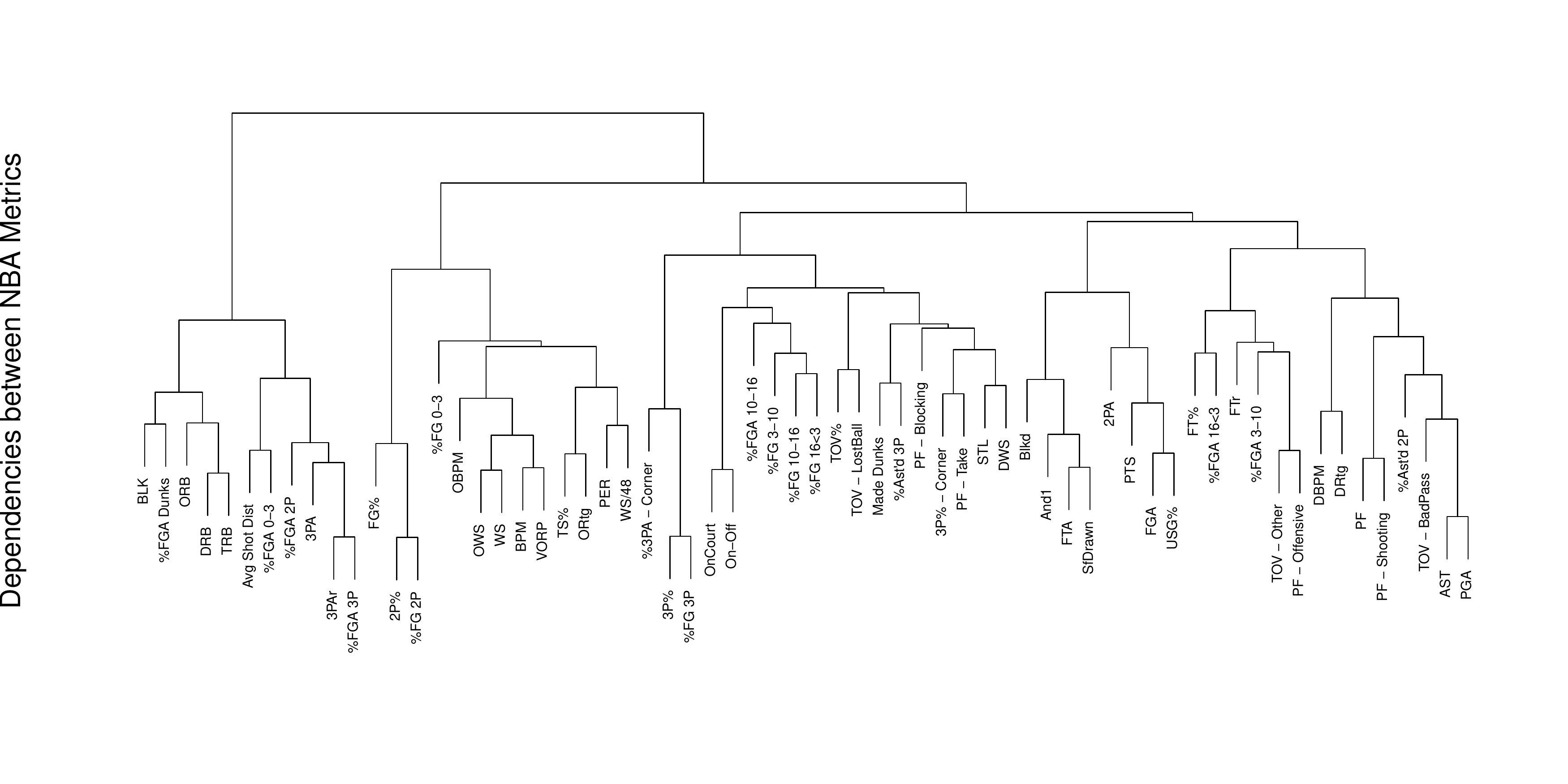}
\end{figure} 

\clearpage

\section*{Proof of $0 \leq \mathcal{S}_m \leq 1$}
We calculate stability for metric $m$ \eqref{eqn:stab} as 
\begin{equation}
\mathcal{S}_m = 1 - \frac{E_m[V_{pm}[X] - V_{spm}[X]]}{V_m[X] - E_m[V_{spm}[X]]}.
\end{equation}
To show $0 \leq \mathcal{S}_m \leq 1$, it suffices to show both
\begin{itemize}
\item[(A)] $E_m[V_{pm}[X] - V_{spm}[X]] \geq 0$
\item[(B)] $V_m[X] - E_m[V_{spm}[X]] - E_m[V_{pm}[X] - V_{spm}[X]] \geq 0$.
\end{itemize}
To verify (A), we can write
\begin{align*}
E_m[V_{pm}[X] - V_{spm}[X]] &=  E_m[V_{pm}[E_{spm}[X]] + E_{pm}[V_{spm}[X]] - V_{spm}[X]]  \\
 & = E_m[V_{pm}[E_{spm}[X]]] + E_m[E_{pm}[V_{spm}[X]]] - E_m[V_{spm}[X]] \\
 & = E_m[V_{pm}[E_{spm}[X]]] \\
 & \geq 0.
\end{align*}
To check (B), note that
\begin{align*}
V_m[X] - E_m[V_{spm}[X]] - E_m[V_{pm}[X] - V_{spm}[X]] &= V_m[X] - E_m[V_{pm}[X]] \\
 &= V_m[E_{pm}[X]] \\
 & \geq 0.
\end{align*}
 
\clearpage

\section*{Glossary of Metrics}

\begin{table}[ht!]
  \renewcommand{\baselinestretch}{1}
\caption{Glossary of NBA metrics used.  All stats are per 36 minutes unless
  otherwise noted.  See \citep{bballref} for more detail.}
\footnotesize
\centering
\begin{tabular}{l|l}
Metric & Description \\ 
  \hline
MP & Minutes played \\ 
  FGA &  Field goal attempts\\ 
  FG\% &  Field goal percentage\\ 
  3PA &  3 point attempts\\ 
  3P\% &  3 point percentage \\ 
  2PA &  2 point attempts\\ 
  2P\% &  2 point percentage\\ 
  FTA &  Free throw attempts\\ 
  FT\% &  Free throw percentage\\ 
  PF &  Personal fouls\\ 
  PTS &  Points\\ 
  PER &  Personal efficiency rating\\ 
  TS\% &  True shooting percentage\\ 
  3PAr &  Three point attempt rate\\ 
  FTr &  Free throw attempt rate\\ 
  ORB &  Offensive rebounds\\ 
  DRB &  Defensive rebounds\\ 
  TRB &  Total rebounds\\ 
  AST &  Assists\\ 
  STL &  Steals\\ 
  BLK &  Blocks\\ 
  TOV\% &  Turnover percentage (per possession)\\ 
  USG\% &  Usage per\\ 
  OWS &  Offensive win shares\\ 
  DWS &  Defensive win shares\\ 
  WS &  Win shares\\ 
  WS/48 &  Win shares per 48 minutes\\ 
  OBPM &  Offensive box plus minus\\ 
  DBPM &  Defensive box plus minus\\ 
  BPM &  Box plus minus\\ 
  VORP &  Value over replacement\\ 
  ORtg &  Offensive rating\\ 
  DRtg &  Defensive rating\\ 
  Avg Shot Dist & Average shot distance \\ 
   \hline
\end{tabular}
\end{table}

\renewcommand{\baselinestretch}{1}
\begin{table}[ht!]
\vspace{-2.2in}
\caption{NBA Glossary cont.}
\centering
\begin{tabular}{l|l}
Metric & Description \\ 
  \hline
  \%FGA 2P &  percentage of field goal attempts that are 2 pointers\\ 
  \%FGA 0-3 &  percentage of field goal attempts within 0-3 feet\\ 
  \%FGA 3-10 &  percentage of field goal attempts within 3-10 feet\\ 
  \%FGA 10-16 &  percentage of field goal attempts within 10-16 feet\\ 
  \%FGA 16$<$3 &  percentage of field goal attempts between 16 feet
                 and the 3 point line\\ 
  \%FGA 3P &  percentage of field goal attempts that are 3 pointers \\ 
  \%FG 2P & percentage of made field goals that are 2 pointers \\ 
  \%FG 0-3 &  percentage of made field goals within 0-3 feet\\ 
  \%FG 3-10 & percentage of made field goals within 3-10 feet \\ 
  \%FG 10-16 &  percentage of made field goals within 10-16 feet\\ 
  \%FG 16$<$3 &  percentage of made field goals between 16 feet and
                the 3 point line\\ 
  \%FG 3P &  percentage of made field goals that are 3 pointers\\ 
  \%Ast'd 2P & percentage of made 2 point field goals that are assisted \\ 
  \%FGA Dunks &  percentage of field goal attempts that are dunks \\ 
  Made Dunks & made dunks (per 36 MP)\\ 
  \%Ast'd 3P &  percentage of made 3 point field goals that are assisted\\ 
  \%3PA - Corner &  percentage of 3 point field goal attempts taken
                   from the corner\\ 
  3P\% - Corner &  3 point field goal percentage from the corner\\ 
  OnCourt &  plus/minus per 100 possessions\\ 
  On-Off &  plus/minus net per 100 possession\\ 
  TOV - BadPass &  turnovers from bad passes\\ 
  TOV - LostBall & turnovers due to lost ball \\ 
  TOV - Other &  all other turnovers (traveling, out of bounds, etc) \\ 
  PF - Shooting &  shooting fouls committed\\ 
  PF - Blocking &  blocking fouls committed \\ 
  PF - Offensive &  offensive fouls committed\\ 
  PF - Take &  take fouls committed\\ 
  PGA &  points generated by assists\\ 
  SfDrawn &  shooting fouls drawn\\ 
  And1 &  shots made on fouls drawn\\ 
  Blkd &  field goal attempts that are blocked\\
\end{tabular}
\end{table}

\begin{table}[ht]
 \renewcommand{\baselinestretch}{1}
\caption{Glossary of hockey metrics used.  All metrics are normalized by total time on
  ice (TOI) unless otherwise noted.}
\centering
\begin{tabular}{l|l}
metric & description \\ 
  \hline
G &  goals\\ 
  A &  assists\\ 
  PTS &  points\\ 
  $\pm$ &  plus / minus\\ 
  PIM &  penalties in minutes\\ 
  EV &  even strength goals\\ 
  PP &  power play goals\\ 
  SH &  short handed goals\\ 
  GW &  game winning goals\\ 
  S & shots on goal \\ 
  S\% &  shooting percentage\\ 
  TSA &  total shots attempted\\ 
  TOI &  time on ice\\ 
  FO\% &  face off win percentage\\ 
  HIT &  hits at even strength\\ 
  BLK &  blocks at even strength\\ 
  TK &  takeways\\ 
  GV &  giveaways\\ 
  GC &  goals created\\ 
  TGF &  total goals for (while player was on the ice)\\ 
  PGF &  power player goals for (while player was on the ice)\\ 
  TGA &  total goals against (while player was on the ice)\\ 
  PGA &  power player goals against (while player was on the ice)\\ 
  OPS &  offensive point shares\\ 
  DPS &  defensive point shares\\ 
  PS &  total point shares\\ 
  CF & Corsi for (on ice shots+blocks+misses)\\ 
  CA &  Corsi against (on ice shots+blocks+misses)\\ 
  CF\% &  Corsi for percentage: CF / (CF + CA)\\ 
  CF\% rel & Relative Corsi for (on ice CF\% - off ice CF\%)  \\ 
  FF &  Fenwick for (shots+blocks+misses)\\ 
  FA &   Fenwick against (shots+blocks+misses)\\ 
  FF\% & Fenwick for percentage: FF / (FF + FA) \\ 
  FF\% rel & Relative Fenwick for (on ice FF\% - off ice FF\%)  \\
  oiSH\% & Team on ice shooting percentage while player on the ice \\ 
  oiSV\% & Team on ice save percentage while player on the ice \\ 
  PDO &  Shooting percentage plus save percentage\\ 
  oZS\% &  percentage of offensive zone starts while on the ice\\ 
  dZS\% &  percentage of defensive zone starts while on the ice\\ 
   \hline
\end{tabular}
\end{table}

\end{document}